\documentclass{emulateapj}
\lefthead{Faltenbacher et al.} 
\righthead{Alignments between central and satellite halos}
\newcommand{\hMsol}{{\,h^{-1}\rm M}_\odot}
\newcommand{\hMpc}{{\,h^{-1}\rm Mpc}}
\newcommand{\hkpc}{{\,h^{-1}\rm kpc}}

\newcommand{\kmsmpc}{\>{\rm km}\,{\rm s}^{-1}\,{\rm Mpc}^{-1}}
\newcommand{\Rvir}{{\,r_{\rm vir}}}

\renewcommand{\vec}[1]{{\mathbf #1}}
\newcommand{\aII}{\langle|\vec{a}_{\rm GCS}\!\cdot\!\vec{r}|\rangle\,}
\newcommand{\axI}{\langle|\vec{a}_{\rm SAT}\!\cdot\!\vec{r}|\rangle\,}

\newcommand{\acI}{\langle|\vec{a}_{\rm GCS}\!\cdot\!\vec{a}_{\rm SAT}|\rangle\,}

\newcommand{\vel}{\langle|\vec{v}  \!\cdot\!\vec{r}|\rangle\,}
\newcommand{\aIv}{\langle|\vec{a}_{\rm SAT}\!\cdot\!\vec{v}|\rangle\,}
\newcommand{\cIv}{\langle|\vec{a}_{\rm GCS}\!\cdot\!\vec{v}|\rangle\,}
\begin{document}
\title{Spatial and kinematic alignments between central and satellite halos} 
\author {A. Faltenbacher\altaffilmark{1},  Y.P. Jing\altaffilmark{1}, 
Cheng Li\altaffilmark{1}, Shude Mao\altaffilmark{2},
H.J. Mo\altaffilmark{3}, Anna Pasquali\altaffilmark{4} and Frank C. van
den Bosch\altaffilmark{4}}
\altaffiltext{1} {Shanghai Astronomical Observatory, Joint Center for
Cosmology and Astrophysics of the Max-Planck-Institut fuer Astrophysik
and the  Shanghai  Astronomical Observatory,  Nandan Road  80,  Shanghai
200030, China}
\altaffiltext{2} {University of Manchester, Jodrell Bank Observatory,
Macclesfield, Cheshire SK11 9DL, UK}    
\altaffiltext{3}{Department of Astronomy, University of Massachusetts,
Amherst MA 01003-9305}
\altaffiltext{4} {Max-Planck-Institute for Astronomy, K\"onigstuhl 17,
D-69117 Heidelberg, Germany }

\begin{abstract}
Based   on a cosmological  N-body  simulation  we analyze  spatial and
kinematic alignments of satellite halos  within  six times the  virial
radius of  group  size host halos ($\Rvir$).   We measure three
different types of  spatial   alignment: halo alignment    between the
orientation   of   the group    central substructure  (GCS)    and the
distribution   of  its   satellites, radial   alignment   between  the
orientation of   a satellite and  the  direction towards  its GCS, and
direct alignment  between the orientation of  the GCS and that  of its
satellites.  In analogy we use the  directions of satellite velocities
and  probe three further   types  of  alignment: the radial   velocity
alignment  between the satellite velocity  and connecting line between
satellite and GCS, the halo velocity alignment between the orientation
of the  GCS and satellite  velocities and  the auto velocity alignment
between the satellites orientations  and  their velocities.  We   find
that satellites are preferentially located along the major axis of the
GCS within at least  $6\Rvir$  (the range probed here).   Furthermore,
satellites preferentially point  towards the GCS.  The most pronounced
signal is detected on small scales but a detectable signal extends out
to $\sim  6\Rvir$.  The direct alignment signal  is weaker,  however a
systematic  trend is   visible  at  distances $\lesssim2\Rvir$.    All
velocity alignments are highly significant on  small scales.  The halo
velocity  alignment is constant   within $2\Rvir$ and declines rapidly
beyond.  The  halo  and the  auto velocity  alignments are maximal  at
small scales and disappear beyond  1 and $1.5\Rvir$ respectively. Our
results suggest that the halo alignment reflects the filamentary large
scale structure which  extends  far  beyond the  virial radii   of the
groups.  In contrast, the main  contribution  to the radial  alignment
arises from the adjustment of the satellite  orientations in the group
tidal  field.   The projected data  reveal good  agreement with recent
results derived from large galaxy surveys.
\end{abstract}
\section{Introduction}
Over  the  last  decades  observational  and  numerical  evidence  has
substantiated the  picture of  a filamentary large-scale  structure in
the universe.  In principle the large-scale tidal field is expected to
induce large-scale correlations between  the orientations of halos and
galaxies     that    are     embedded    within     these    filaments
\citep[e.g.,][]{2000ApJ...543L.107P,2000ApJ...545..561C,
2000MNRAS.319..649H,         2001MNRAS.320L...7C,2001ApJ...559..552C,
2002MNRAS.332..339P, 2002MNRAS.335L..89J}.   On the other  hand, the
subsequent accretion onto larger  systems, such as groups and clusters
of galaxies,  may alter the orientations of  these (sub-)structures in
response   to  the   local  tidal   field  \citep{1994MNRAS.270..390C,
2005ApJ...629L...5L}.   Cosmological  N-body  simulations provide  a
valuable  tool  to  differentiate  the various  contributions  to  the
halo/galaxy alignments within overdense regions.

Observationally, various types of alignment between galaxies and their
environment have  been detected  on  a   wide  range in  scales,  from
super-cluster systems   down  to  the distribution of   the  satellite
galaxies in our  Milky Way (MW).   On cluster scales various different
types of alignment are discussed in the literature : alignment between
neighboring clusters \citep{1982A&A...107..338B,  1989ApJ...338..711U,
1989ApJ...344..535W,     1994ApJS...95..401P,    2002ApJ...565..849C},
between brightest cluster galaxies (BCGs) and their parent clusters
\citep{1980MNRAS.191..325C,  1982A&A...107..338B, 1990AJ.....99..743S,
1991A&A...243...38R,2002MNRAS.329L..47P}, between the orientation of
satellite    galaxies   and   the    orientation   of    the   cluster
\citep{1985ApJ...298..461D,  2003ApJ...594..144P},   and  between  the
orientation  of satellite  galaxies  and the  orientation  of the  BCG
\citep{1990AJ.....99..743S}.  According  to these studies  the typical
scales  over which  clusters reveal  signs for  alignment range  up to
$10-50\hMpc$, which can be most naturally explained by the presence of
filaments.

With  large galaxy  redshift  surveys, such  as  the two-degree  Field
Galaxy Redshift  Survey \citep[2dFGRS,][]{2001MNRAS.328.1039C} and the
Sloan Digital Sky  Survey \citep[SDSS,][]{2000AJ....120.1579Y}, it has
recently  also  become possible  to  investigate  alignments on  group
scales  using large  and homogeneous  samples.  This  has  resulted in
robust  detections of various  alignments: \cite{2005ApJ...628L.101B},
\cite{2006MNRAS.369.1293Y}  and  \cite{2007MNRAS.376L..43A} all  found
that satellite galaxies are preferentially distributed along the major
axes  of  their host  galaxies,  while \cite{2005ApJ...627L..21P}  and
\cite{2006ApJ...644L..25A} noticed that  satellite galaxies tend to be
oriented towards the galaxy at the center of the halo.

In   contradiction to   the  studies above, \cite{1969ArA.....5..305H}
found    that    satellites   around   isolated   late  type  galaxies
preferentially  lie  along the  minor  axis  of the  disc.  Subsequent
studies, however, were   unable  to confirm this so-called   `Holmberg
effect'\citep{1975AJ.....80..477H,                1979MNRAS.187..287S,
1982MNRAS.198..605M,          1997ApJ...478L..53Z}.           Recently
\cite{2007arXiv0704.3441A} reported    a  Holmberg effect    at  large
projected distances around blue host galaxies, while on smaller scales
the satellites were found to  be aligned with the  major axis of their
host  galaxy  and   \cite{2005ApJ...627..647B} claim that   a  careful
selection of isolated late-type galaxies  reveals the the tendency for
the  satellites to    align  with the   minor  axis  of  the  galactic
disc.  Investigating  the companions of M31 \cite{2006AJ....131.1405K}
find little evidence  for a Holmberg effect.  Yet,  the Milky Way (MW)
seems to exhibit a  Holmberg effect even on small  scales, in that the
11  innermost MW satellites  show   a pronounced planar   distribution
oriented       close   to    perpendicular     to     the    MW   disc
\citep{1982Obs...102..202L,  1994ApJ...431L..17M, 2005A&A...431..517K,
2005A&A...437..383K, 2005MNRAS.363..146L}.

Numerical  simulations  have been employed  to    test alignment on  a
similar range in scales, from  super-clusters down to galaxy-satellite
systems.   All  studies  focusing on   cluster size  halos   report  a
correlation in  the orientations for distances  of at least $10\hMpc$;
some studies observe a positive alignment signal up to $100\hMpc$
\citep[e.g.,][]{2000MNRAS.319..614O,2002A&A...395....1F,
2005MNRAS.362.1099F,2005ApJ...618....1H,2005ApJ...629..781K,
2006MNRAS.365..539B}.  These findings are interpreted as the signature
of the filamentary   network which  interconnects  the  clusters.  The
preferential accretion along  these  filaments causes the  clusters to
point  towards each other.  Also,  for galaxy and  group-sized halos a
tendency to point toward neighboring halos is detected.  According to
\cite{2006MNRAS.370.1422A} the  alignments for such  intermediate mass
objects are  caused by  tidal fields  rather than  accretion along the
filaments.  Consequently, the mechanisms responsible for the alignment
of the orientations depend  on halo mass.  Further evidence for a mass
dependence   of alignment effects comes   from  the examination of the
halos' angular momenta.
\cite{2005ApJ...627..647B} and \cite{2007ApJ...655L...5A} find that the
spins of  galaxy  size  halos  tend to be   parallel  to the filaments
whereas the spins of group-sized halos tend  to be perpendicular. This
behavior may originate in the relative sizes  of halos with respect to
the surrounding filaments.  

On subhalo scales basically  three different types of  alignments have
been discussed: the alignment of the overall subhalo distribution with
the orientation of  the host halo \citep[e.g.,][]{2004ApJ...603....7K,
2005ApJ...629..219Z,      2006ApJ...650..550A,    2007MNRAS.378.1531K,
2007MNRAS.374...16L},  the alignment  of the  orientations of subhalos
among  each    other  \citep[e.g.,][]{2005ApJ...629L...5L}  and,  very
recently, the orientation of the satellites with respect to the center
of the host  \citep{2007arXiv0705.2037K,2007arXiv0707.1702P}.   Again,
accretion along the filaments and the impact of tidal fields have been
invoked as explanations  for the former  and the latter, respectively.
Thus, on  all scales tidal   fields and accretion along filaments  are
considered to be  the  main  contributers  to the observed   alignment
signals.  Here we attempt to isolate  the different contributions.  In
particular we focus on the continuous  transition from subhalo to halo
scales meaning we examine the  alignment of (sub)structure on distance
scales between 0.3  and 6  times the virial   of  groups sized  halos.

Faltenbacher et al. (2007, hereafter Paper~I)\nocite{2007ApJ...662L..71F} 
applied the halo-based  group finder  of \cite{2005MNRAS.356.1293Y} to
the  SDSS  Data Release  Four \citep[DR4;][]{2006ApJS..162...38A}  and
carried out a study of  the mutual alignments between central galaxies
(BCG) and their satellites in group-sized halos. Using the same
data set consisting of over  $60000$ galaxies three different types of
alignment  have been investigated : (1)   the `halo' alignment between
the orientations  of the  BCG  and associated  satellite distribution;
(2) the  `radial' alignment between  the direction  given  by the
BCG-satellite connection   line and  the satellite  orientation;  (3)
the `direct' alignment between the orientations of the BCG and the
satellites.  The study presented in  this paper focuses on  the same
types of alignment and  is aimed  to compare  the observational
results with theoretical expectations derived from N-body simulations.

There are a variety of dynamical processes which can contribute to the
alignments  of satellites associated  with groups,  the most important
are:  (1) a possible   pre-adjustment of satellites  in the filaments,
which for distances of  a few times  the virial radius commonly  point
radially towards the group; (2) the preferential accretion along those
filaments; (3)  the  change of the  satellite  orbits in the  triaxial
group potential well;  (4) the  continuous re-adjustment of  satellite
orientations as they orbit within the  group. Basically, the first two
points can be attributed to the large  scale environment of the groups
whereas the latter two are more closely associated  with the impact of
the  group potential  on  small scales.   The purpose   of the present
analysis is to separate the   different contributions to the  observed
alignment signals, therefore   we analyse the mutual orientations   of
satellites  within 6 times the virial  radius of the groups. 
Since the tidal forces are  closely related to   the dynamics of  the
satellites additional insight into  the generation of alignment can be
gained by considering  the  satellite velocities.  Therefore, we  also
investigate the direction of the satellite  velocities with respect to
their   orientations, which constitutes an   indirect way to infer the
impact of the dynamics onto the orientation of the satellites.  A more
direct  way to  work out the  interplay between  the dynamics  and the
orientations would  be to trace  the orbits  of individual satellites,
however such an approach  goes beyond the  scope  of the present
study.

The  paper is organized as  follows.  In \S~\ref{sec:sim} we introduce
the    simulation       and       describe     the     halo    finding
procedure. \S~\ref{sec:size} deals with some technical aspects, namely
the   determination  of the  size     and  	orientation  of    the
substructures.  In \S~\ref{sec:3D} we present the signals of the three
dimensional spatial and velocity alignments  and in \S~\ref{sec:2D} we
repeat the analysis based on projected data. Finally, we conclude with
a summary in
\S~\ref{sec:diss}. 

\section{Simulation and halo identification}
\label{sec:sim}
For the present analysis  we employ an  N-body simulation of structure
formation in   a flat $\Lambda$CDM   universe with   a matter  density
$\Omega_m=0.3$, a  Hubble parameter  $h=H_0/(100 \kmsmpc)=0.7$,  and a
Harrison-Zeldovich    initial power   spectrum  with     normalization
$\sigma_8=0.9$.  The  density  field is sampled by   $512^3$ particles
within a $100\hMpc$ cube resulting in  a mass resolution of $6.2\times
10^8\hMsol$.   The softening length was  set  to $\epsilon = 10\hkpc$,
beyond which the gravitational force between  two particles is exactly
Newtonian. The density  filed is evolved with 5000  time steps from an
initial   redshift  of $z_i=72$  using  a  PPPM  method.  An extensive
description of the simulation can be found in
\cite{2002ApJ...574..538J} where it is quoted as LCDMa realization.  

As detailed in the following two paragraphs the host halos and its
satellites are found in two subsequent steps with two different
techniques, first the main halos are located thereafter the associated
satellite halos are detected. In  order to   identify  the host halos
we  first  run   a FoF   algorithm \citep{1985ApJ...292..371D} on  
the simulation output at $z=0$.  We set the  FoF linking length to 0.1
times  the mean particle  separation, which  selects  regions with  an
average overdensity  of $\sim3000$. Note that, this linking length is a
factor of two smaller than the commonly used value of 0.2,
consequently only the central part of the host halo (and occasionally
large substructures) are selected.   
Subsequently, the virial radius,
$\Rvir$, is  defined as the radius of  the sphere centered on the most
bound FoF  particle  which includes a  mean  density of  101 times the
critical density, and we simply define the virial mass of each halo as
the mass within $\Rvir$. If the virial  regions of two halos overlap,
the  lower mass halo is discarded.   In what follows  we only focus on
the 515 halos with  a virial mass in the  range from $10^{13} \hMsol$
to $5\times10^{14}\hMsol$  (corresponding  to  halos  with  more than
16,000 particles).  Since  this is  the  typical mass scale of  galaxy
groups, we will refer to these halos as `groups'.

In a second step we search for self-bound (sub)structures using the SKID
halo finder \citep{2001PhDT........21S} applied to the particle 
distribution within group centric distances of $6\Rvir$.   As  discussed  in 
\cite{2006MNRAS.366.1529M}  SKID  adequately  identifies the  smallest
resolvable  substructures  when using a  linking length   $l$ equal to
twice the  softening  length, i.e.   four  times the spline  softening
length.   We    therefore adopt  $l=20\hkpc$.    Throughout  we   will
distinguish between  ``group central substructures'' (GCSs), which are
located at the center of our groups, and satellites  which are all the
other (sub)structures, no   matter whether they  lie within  or beyond
$\Rvir$.  According to this definition every group hosts one, and only
one, GCS at  its center while it  may have numerous satellites outside
the  volume occupied by  the   GCS.  Satellites  are allocated to  all
groups from  which they  are separated less   than $6\Rvir$. Hence,  a
satellite may be assigned to more than one GCS.
\section{Size and orientation of substructures}
\label{sec:size}
Before describing the computation  of the orientation we determine the
typical   sizes of the GCSs  and   the satellites. Knowledge about the
physical sizes of the (sub)structures provides a crucial link for the
comparison to observational data.
\subsection{Sizes of  group central substructures} 
\begin{figure}
\plotone{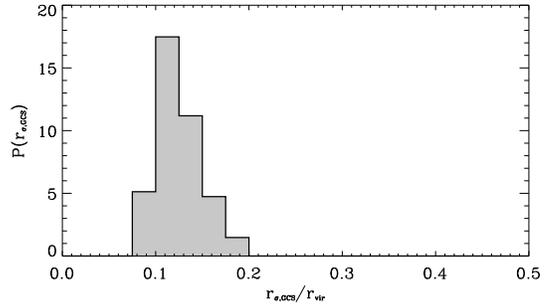}
\caption{\label{fig:covi}
Distribution of the spatial dispersion $r_\sigma$ of the group central
substructure (GCS) in units of  the virial radius. Satellites can only
be resolved at radii  larger  than the size of the GCS.}
\end{figure}
The physical   interpretation    of the  size  of   the   GCS  is  not
straightforward. For one thing, it  depends on the SKID linking length
used.  However, for our  purposes it is sufficient  to notice that the
GCS represents the dense inner region of  the group which, largely due
to numerical reasons,    is free of substructure.   Consequently,  any
radial dependence  of satellite properties can only  be probed down to
the size of the GCS. In order to express the  sizes of the GCS and the
satellites we   use the  rms   of  the  distances between   the  bound
particles, $r_{\sigma}$.  The  advantage of this  size measure is that
it provides a direct  estimate of the  (momentary) size without having
to make any assumption regarding the  actual density distribution.  In
the case of an isolated NFW  halo $r_\sigma\approx0.5\Rvir$, with only
a  very    weak    dependence    on  the     concentration  parameter.
Figure~\ref{fig:covi} displays the $r_\sigma$  distribution of the  GCSs
in units   of the group's  virial  radius, $\Rvir$.   The distribution
peaks at $0.11\Rvir$ and has a mean of $0.13\Rvir$.
\subsection{Sizes of satellite halos} 
\begin{figure}
\plotone{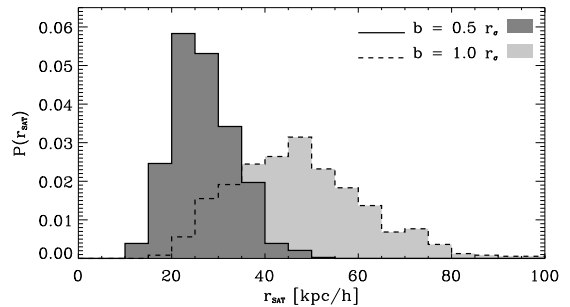}
\caption{\label{fig:rkpc}
Distribution of the radii of satellites found within the virial radius
of the  group.   In  this context radius    refers to listed  fraction
(0.5, 1.0)  of the  satellites  spatial dispersion $r_\sigma$.  For
example, the typical inner radii  probed by the $b=0.5r_\sigma$ sample
is $\sim30\hkpc$.}
\end{figure}
The aim of the  present analysis is  twofold: (1) to assess the impact
of the  group tidal field  on the  satellite  orientations, and (2) to
compare the alignment signals in our N-body simulation to observations
of galaxy alignments.  The impact of the group tidal field is stronger
at larger satellite-centric radii.  On the  other hand, since galaxies
reside at the centers of their dark matter halos, the central parts of
the  satellites   are more of   interest  when comparing the alignment
signals with those observed for galaxies. To meet both requirements we
therefore measure the  orientation of the  satellite mass distribution
within  two radii.   In analogy to   the measurement of GCS  sizes, we
determine these   radii  with reference   to  the spatial   dispersion
$r_\sigma$.   More precisely, we choose  the particles  within 1.0 and
$0.5r_\sigma$  as the basic sets for   the subsequent determination of
the satellite     orientation   (see  Section~\ref{sec:ori}    below).
Figure~\ref{fig:rkpc} displays  the distributions of the corresponding
physical   sizes.   The  $0.5r_\sigma$    sample  probes the    matter
distribution of   the    satellites within  $\sim25\hkpc$,   which  is
comparable  to the sizes of elliptical   galaxies.  The mean, physical
radii  of  the $1.0r_\sigma$  sample is  $\sim50\hkpc$.  If not quoted
otherwise  we will display the results  for  the $0.5r_\sigma$ sample,
since this may most    closely resemble the properties of   observable
galaxy distributions (outside  of the  very  central part of the  host
halo).
\subsection{Orientation} 
\label{sec:ori}
\begin{figure}
\plotone{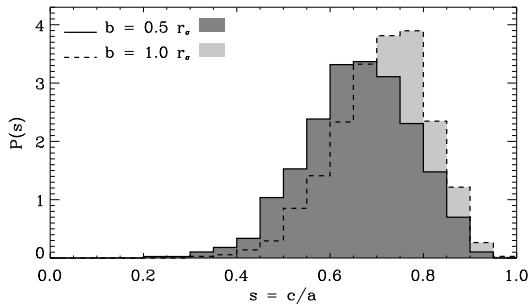}
\caption{\label{fig:shape}
Distribution of satellite shapes, represented by the ratio of shortest
to the  longest semi-major axis of   the mass-ellipsoid ($s=c/a$).  The
colors  correspond to the samples with  different  truncation radii as
listed. With  increasing  size    the   halos become  rounder.     The
distribution of    the  $b=0.5r_\sigma$ sample  is   rather symmetric,
whereas for larger truncation radii there appears a slight asymmetry.}
\end{figure}
There   are   a  few   different   ways   found   in  the   literature
\citep[e.g.,][]{2002sgdh.conf..109B,               2002ApJ...574..538J,
2005ApJ...627..647B,  2005ApJ...629..781K,  2006MNRAS.367.1781A}  to
model  halos as  ellipsoids.  They  all differ  in details,  but most
methods  model halos  using the  eigenvectors from  some form  of the
inertia tensor.   The eigenvectors correspond to the  direction of the
major axes, and the eigenvalues  to the lengths of the semi-major axes
$a\geq b  \geq c$.  Following  \cite{2006MNRAS.367.1781A} we determine
the main  axes by iteratively  computing the eigenvectors of  the {\it
distance weighted} inertia tensor.
\begin{equation}
I_{ij} = \sum_{k=1,N} {r_{ki}r_{kj}\over r^2_{k}}\ , 
\end{equation}
where $r_{ki}$ denotes  the $i$th component  of the position vector of
the $k$th particle with respect to the center of mass and
\begin{equation}
\label{eq:weight}
r_k = \sqrt{{x^2\over a^2} + {y^2\over b^2} + {z^2\over c^2}}
\end{equation}
is the elliptical distance  in the eigenvector coordinate  system from
the center to the $k$th particle.  The square roots of the eigenvalues
of the inertia tensor determine the axial ratios of the halo ($a:b:c =
\sqrt{\lambda_a}:\sqrt{\lambda_b}:\sqrt{\lambda_c}$). The iteration is 
initialized by computing the eigenvalues of the inertia tensor for the
spherically truncated halo.  In the following iterations the length of
the intermediate axis is kept unchanged and all bound particles within
the ellipsoidal window determined by  the eigenvalues of the foregoing
iteration are used for the computation of the new inertia tensor.  The
iteration is completed   when the eigenvectors    have converged.  The
direction   of    the resulting  major axis    is   identified  as the
orientation.  The advantage of keeping  the intermediate axis fixed is
that  the number of  particles within the  varying ellipsoidal windows
remains almost  constant.  Instead, if  the longest (shortest) axis is
kept constant the  number of particles  within the ellipsoidal windows
can decrease (increase) substantially during the iteration.

Note  that we apply this   truncation   to all (sub)structures,   both
satellites and  GCSs, and that  the orientation of each sub(structure)
is measured   within   this truncation  radius.   Throughout  we  only
consider those sub(structures)   that   comprise at least 200    bound
particles within the volume of the final ellipsoid (corresponding to a
lower limit  in mass of  $\approx 10^{11}\hMsol$).  For the satellites
this implies that a   smaller truncation radius  results in  a smaller
sample.  For example,   there are 772 $0.5r_\sigma$ satellites  within
the virial radii   of  our groups   whereas  the $1.0r_\sigma$  sample
comprises 1431 satellites.  Since all 515 GCSs  contain more than 200
particles within $0.5r_\sigma$ their sample size is independent of the
truncation radius used.

Figure~\ref{fig:shape}  displays the    distribution  of   the   shape
parameter $s=c/a$.   The  shading corresponds to  different truncation
radii as  listed.  There is a  weak  indication that satellites become
more spherical with  increasing truncation radii.  A  similar behavior
was found   for  isolated halos   \citep[e.g.,][]{2002ApJ...574..538J,
2006MNRAS.367.1781A}.  As  discussed by \cite{2006MNRAS.367.1781A} the
exact determination   of individual shapes   may need as  many as 7000
particles, so  that  the resolution  of the present  simulation is not
suited for  the analysis of  (sub)structure shapes.   However, for the
determination of the orientations, which is the focus of this paper, a
particle limit of 200 can be considered conservative
\citep[cf.,][]{2002MNRAS.335L..89J, 2007arXiv0707.1702P}.   A study
examining the shapes of substructure in  a single high-resolution
Milky  Way-sized halo can be found in \cite{2007arXiv0705.2037K}.
\section{Three dimensional Alignments}
\label{sec:3D}
For both classes of objects, GCSs and satellites, the orientations are
determined according to    the  approach described  above.    A  third
orientation-like quantity is   given by  the   direction of the   line
connecting a    GCS-satellite  pair.   Throughout  we  refer   to  the
orientation  of  the GCS,  the satellite and   the  connecting line as
$\vec{a}_{\rm  GCS}$, $\vec{a}_{\rm SAT}$ and $\vec{r}$, respectively.
These quantities are unit vectors, such that the scalar product of two
vectors yields the cosine of the angle between them.  We will focus on
three   different types of alignment,   (1)  the {\it halo  alignment}
between the orientations of the GCSs and the connecting lines, (2) the
{\it radial alignment} between the  orientations of the satellites and
the  connecting lines and (3) the  {\it  direct alignment} between the
orientation of the  GCS and that  of its satellites.  In  addition, we
also consider   various   alignments based   on  the  proper velocity,
$\vec{v}$, of  the satellite with respect to  its GCS.  In particular,
we  discuss (4) the  {\it radial velocity alignment} between $\vec{v}$
and  $\vec{r}$,   (5)  the  {\it   halo   velocity alignment}  between
$\vec{a}_{\rm GCS}$  and $\vec{v}$,  and finally  (vi) the  {\it  auto
velocity alignment} between the orientations, $\vec{a}_{\rm SAT}$, and
velocities, $\vec{v}$, of the  satellites.  Here $\vec{v}$ is the unit
vector indicating  the direction of the  {\it proper}  velocity of the
satellite (including the Hubble flow) relative to the host.  Since all
the  other quantities also represent  unit vectors the scalar products
yield the cosines of the enclosed angles.
\subsection{Halo alignment}
\label{sec:halo3D}
\begin{figure}
\plotone{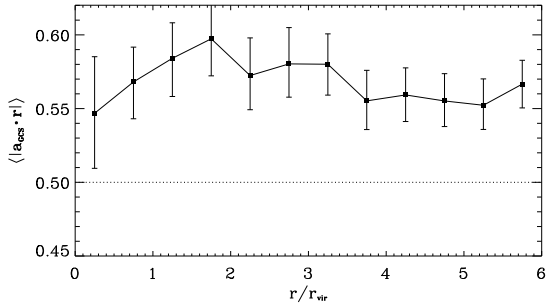}
\caption{\label{fig:a11}
Mean values of  the cosines of  the angles between the orientations of
the GCS  and  the connecting lines   to the satellites,  $\aII$,  as a
function  of  $r/\Rvir$   for the   $0.5r_\sigma$ sample.  The  dotted
horizontal   line   indicates   the mean  values     for an  isotropic
distribution. The error bars  indicate the $95\%$ bootstrap confidence
intervals within each distance bin.}
\end{figure}

In  order to measure the  alignment between the  GCS and the satellite
distribution we use $\vec{a}_{\rm GCS}$ and $\vec{r}$ (the orientation
of the GCS and the position of the satellite with respect to its GCS).
Figure~\ref{fig:a11} displays the  radial dependence  of $\aII$ within
$6\Rvir$, where $\langle\cdot\rangle$  denotes the mean value within a
bin of  $r/\Rvir$. The   error  bars  indicate the   $95\%$  bootstrap
confidence intervals based on 1000 bootstrap samples for each distance
bin.
Over the  entire  range of distances   probed, the mean  values of the
cosines deviate  significantly from   a isotropic  distribution.   The
strength of    the alignment, i.e.    the  deviation  from $\aII=0.5$,
increases  with group   centric distance  and   reaches  a maximum  at
$\sim1.7\Rvir$.  The subsequent  decline,  however, is very  weak  and
even at   $6\Rvir$     the alignment   is     still very    pronounced
($\aII\approx0.55$), with no  clear indication  of a  downward  trend.
The  fact  that there  is  strong  alignment over  such   a long range
suggests that the halo intrinsic alignment is closely connected to the
filamentary structure in which the groups are embedded in.  Since here
we focus  on the transition  between  group and environment  dominated
areas  we do  not  aim  to map out   the  entire range of  the  radial
alignment.  

The weakening of  the signal at small scales may  be attributed to the
fact that the information about  the filamentary origin is washed away
once  the satellites  start to  orbit  within the  groups (i.e.,  once
non-linear effects kick in). Yet,  the orientation of the group itself
is  closely correlated  with the  surrounding filamentary  network, so
that a residual alignment is maintained by the overall distribution of
satellites   orbiting   in   the   potential   well   of   the   group
\citep[cf.][]{1987ApJ...321..113S,                 2005ApJ...629..219Z,
2007MNRAS.378.1531K}. Additionally,  if one assumes that filaments are
approximately cylindrical  in shape and  the  GCS is  aligned with the
orientation  of  the  cylinder,   then the   mean  angles between  the
orientation of  the GCS and  the satellites position become  larger at
smaller group-centric radii. In  fact,  at distances smaller than  the
radius     of  the   cylinder    the  distribution   will  converge to
isotropic. Finally, some contribution to the decrease of the alignment
strength on small  scales may come   from the fact  that satellites on
nearly radial orbits are filtered  out during their epicenter passage.
They get severely  stripped and consequently  the  number of particles
that  remains bound can    easily fall below the detection   criterion
(minimum of 200 particles), thus weakening the alignment signal.
\subsection{Radial alignment}
\label{sec:radial3D}
\begin{figure}
\plotone{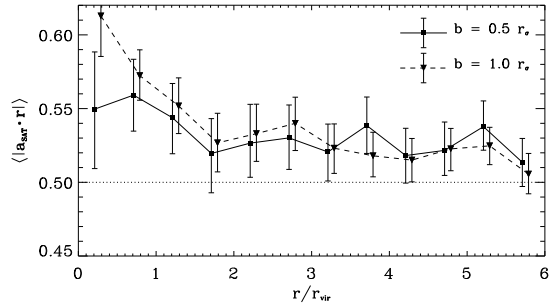}
\caption{\label{fig:ax1}
Same  as  Figure~\ref{fig:a11},  but  for  the  distributions of  cosines
between the satellite orientation and the  connecting line to the GCS,
$|\vec{a}_{\rm SAT}\!\cdot\!\vec{r}|$ for the $0.5$ and
$1.0r_\sigma$ samples.}
\end{figure}
The radial alignment, $\axI$, probes   the orientations of  individual
satellites,  $\vec{a}_{\rm SAT}$,  relative  to the direction pointing
towards their  GCS,  $\vec{r}$.  Figure~\ref{fig:ax1}  displays $\axI$
for  distances  up to  $6\Rvir$.  The  line styles represent different
truncation   radii of the  satellites.    Over   the entire range   of
group-centric   distances probed,   the  data  reveal  a   significant
anisotropic distribution.  The  signal   is most pronounced on   small
scales, where it also  shows  a strong  dependence on the   truncation
radii.   The $1.0r_\sigma$ sample,  which includes the behavior of the
outer   mass shells of  the  satellites,  clearly exhibits a  stronger
deviation from isotropy.  Within $\sim 1.5\Rvir$ there is a pronounced
decline of  the radial alignment signal,  while it  remains remarkably
constant   at larger radii.     For   distance in the  range   between
$2-6\Rvir$ we detect a weak but significant signal, $\acI\approx0.52$,
inconsistent with isotropy at 95\%  confidence level in good agreement
with
\cite{2007arXiv0704.2595H}.           In      a     recent      study,
\cite{2007arXiv0705.2037K}  detected no radial alignment for distances
$\gtrsim3\Rvir$.  However, their analysis  is based on a  resimulation
of  a  single galaxy-sized  host  halo.   Since  this halo   is rather
isolated, in that  it  has  not experienced  any major   merger  after
redshift $z=1.7$, it is likely  that its ambient   filaments have already  been
drained.

At large distances  satellites preferentially reside  in filaments (as
discussed in the context of Figure~\ref{fig:a11}) which point radially
towards    the    groups.    Consequently, the   signal     on  scales
$\gtrsim2\Rvir$ indicates   an   alignment   between  the    satellite
orientations  and the filaments in which  they are  embedded.  Such an
alignment may  be caused by accretion  of matter along those filaments
or by the local tidal fields generated by the mass distribution within
the filaments.  The group tidal field  is not likely to be responsible
for the observed large scale alignment signal due to its rapid decline
with   distance.   On small scales,    however,  the tidal  field  can
substantially  alter the orientations of  the satellites.  As shown by
\cite{1994MNRAS.270..390C} the time scale on which a prolate satellite
can adjust its  orientation to  the  tidal field  of  a group is  much
shorter than the Hubble time, but  longer than its intrinsic dynamical
time.   Therefore,   the  adjustment of   the  satellite  orientations
parallel to the  gradients of the group  potential offers a convincing
explanation for   a radial alignment  signal   on small scales.   This
perception  is  further supported by the   dependence of the alignment
strength on the  truncation radii of  the satellites.  For the largest
radii, which are  strongest  affected by  tidal forces,  the alignment
signal is strongest.
\subsection{Direct alignment}
\begin{figure}
\plotone{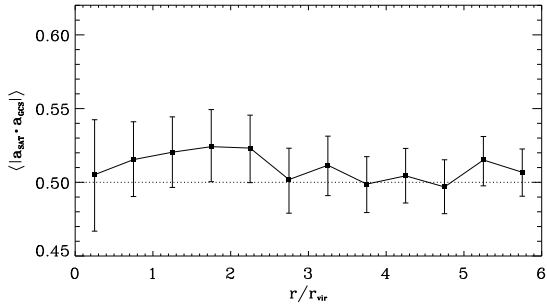}
\caption{\label{fig:ac1}
Same as Figure~\ref{fig:a11},  but for the distributions of
cosines between the orientation of the satellites and the
GCS, $|\vec{a}_{\rm GCS}\!\cdot\!\vec{a}_{\rm SAT}|$.}
\end{figure}
The  strong signals for  halo and  radial alignment   may lead  to the
expectation of a comparably pronounced signal for the direct alignment
between the  orientation of  the   GCS, $\vec{a}_{\rm GCS}$, and   the
orientations of its satellites,  $\vec{a}_{\rm SAT}$.  However, as can
be seen in Figure~\ref{fig:ac1}, the signal  is weak.  There is only a
weak trend for positive  alignment  up to $2\Rvir$.  The  significance
found at  distances between $1$ and $\sim2\Rvir$  seems to be somewhat
higher  ($\sim90\%$   confidence). Based    on an   analytical   model
\cite{2005ApJ...629L...5L}  predict   a   certain degree   of parallel
alignment between host and satellite orientations due to the evolution
of the satellites within the tidal shear field of host. The signal for
the direct  alignment may be  a  relic of  this effect.
 
To  summarize, we find positive alignment  signals for all three types
of alignment tested here.  However, they differ in strength and radial
extent. The halo alignment is the strongest and reaches far beyond the
virial radii of the groups ($\gtrsim6\Rvir$).  The radial alignment is
most pronounced at small scales, where it  reveals a strong dependence
on the radial  extent of the  satellite over which its orientation has
been measured. Although the radial alignment is weak beyond $\sim 1.5
\Rvir$,  the signal  stays  remarkably  constant out  to  $\sim6\Rvir$.   
Finally, the  least prominent signal  comes from the direct alignment.
This ranking of the alignment strengths is  in good agreement with the
observational results reported  in Paper~I.  
\subsection{Alignments based on subhalo velocities}
\label{sec:vel}
If tidal forces give rise to the radial alignment  on small scales, as
displayed in Fig.~\ref{fig:ax1}, the  satellite orientations should be
related to their actual velocities and the local gradients of the host
potential.  For instance a satellite  moving radially towards the  GCS
will show  an  enhanced radial alignment  since  the  gradient of  the
potential and the  actual velocity are pointing  in the same direction
inducing  an orientation in radial direction.   On the other side the
orientations of satellites  moving perpendicular to  the gravitational
field (i.e.  tangentially with respect to the GCS) will lie in between
their velocities and  the gradients of the   potential well.  To  gain
some  more insight  into the  dynamical  origin of  the alignments, we
include  the directions  of  satellite velocities  into  the alignment
study.  We will  consider  three different  kinds  of alignments:  the
radial velocity alignment,  $|\vec{v}\cdot\vec{r}|$, the halo velocity
alignment  $|\vec{a}_{\rm  GCS}\cdot\vec{v}|$  and the  auto  velocity
alignment $|\vec{a}_{\rm SAT}\cdot\vec{v}|$.
\begin{figure}
\plotone{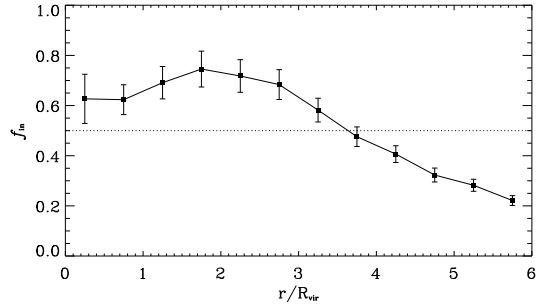}
\caption{\label{fig:io} Ratio of inward moving satellites,
$f_{\rm in} = n_{\rm in}/ (n_{\rm out}  + n_{\rm in})$ for the
$0.5r_\sigma$ sample. Error bars are Poisson.}
\end{figure}
To facilitate the interpretation  of the velocity alignments, we split
the  satellites  according  to  whether  their net  motion  is  inward
($\vec{v} \! \cdot  \! \vec{r} < 0$) or outward ($\vec{v}  \! \cdot \! 
\vec{r} >  0$) with respect  to their group.   Figure~\ref{fig:io} shows
the fraction of inward moving  satellites, $f_{\rm in}$, as a function
of their  group centric distances.   Note that $f_{\rm in}$  reaches a
maximum around  $\sim 2\Rvir$, beyond which the  Hubble flow gradually
starts to  become more and  more important.  In fact,  at sufficiently
large radii, where the Hubble flow dominates, one expects that $f_{\rm
in}=0$, and all satellites reveal an outward motion.  For satellites
that are in virial equilibrium within the group potential (i.e., at $r
\la \Rvir$), one  expects roughly equal numbers of  inward and outward
moving  systems (i.e.,  $f_{\rm  in}=0.5$).  However,  on these  small
scales  one has  an  additional contribution  from  the infall  region
around  the  group,  causing  $f_{\rm  in} >  0.5$.   In  addition,  a
substantial fraction  of satellites  get stripped below  the detection
limit (200 particles)  at their peri-centric passage, so  that they no
longer   contribute   to   the   signal  on   their   outward   motion
\citep[cf.,][]{2007MNRAS.375..313F}.    At   $\Rvir$,   the   outgoing
satellite  fraction  is  about   40\%,  which  is  (within  the  errors)
consistent    with    the   value    $\sim    30\%$   determined    by
\cite{2005MNRAS.364..424W}.  
If one assumes an average ratio of 6:1 between apo- and peri-center distances
for     typical    satellite     orbits    \citep{1998MNRAS.300..146G,
1999ApJ...515...50V} the majority of these satellites must have passed the
central parts of the group before \citep[cf.,][]{2007astro.ph..3337D}.

\begin{figure}
\plotone{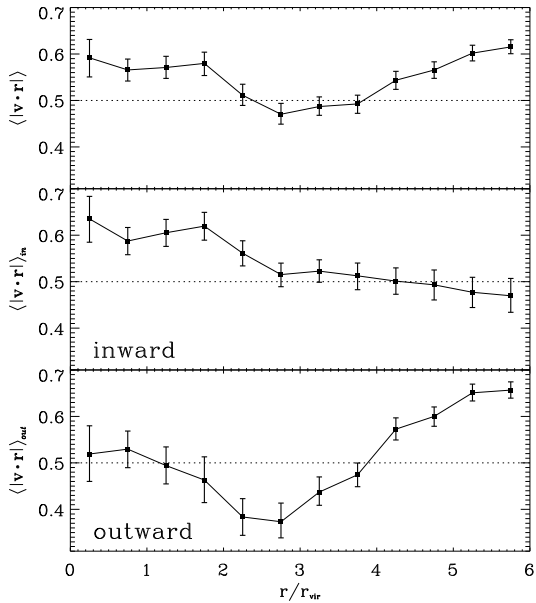}
\caption{\label{fig:vel}
Same as Figure~\ref{fig:a11},  but  for $\vel$.  The  upper, middle  and
lower panel displays the signal  for all, the  inward and the  outward
moving satellites, respectively.}
\end{figure}
The upper panel of  Figure~\ref{fig:vel} displays the radial  velocity
alignment, $\vel$, as a function of $r/\Rvir$.  $\vel > 0.5$ indicates
that the distribution of angles between $\vec{r}$ and $\vec{v}$ is not
isotropic, instead, on   average they preferentially   point in radial
directions. This behavior is in agreement  with earlier studies of the
velocity anisotropy of subhalos  which   is usually expressed  by  the
anisotropy parameter $\beta=1-0.5(\sigma_t/\sigma_r)^2$
\citep[e.g.,][]{1987gady.book.....B}, where  $\sigma_t$ and $\sigma_r$
denote the  velocity dispersions of the  satellites  in the tangential
and radial  direction, respectively.  Note,  $\vel$ is closely related
to  $\beta$.  If one assumes a   relaxed (steady-state) halo the above
mentioned tendency towards  radial  motions translates  into a  higher
radial velocity dispersion compared to the tangential one $\sigma_r >
\sigma_{t1} = \sigma_t/\sqrt{2}$ (where $\sigma_{t1}$ and $\sigma_t$ 
are the  one  and  two  dimensional tangential  velocity  dispersions,
respectively  and tangential isotropy is assumed).   Thus $\vel > 0.5$
on    small  scales   ($r  \la  2\Rvir$)   suggest    that $\sigma_r  >
\sigma_t/\sqrt{2}$,  in  good  qualitative  agreement  with  numerical
simulations which have shown that $\beta >  0$ for subhalos within the
virial radius of their hosts.
\citep{1998MNRAS.300..146G, 2000ApJ...539..561C, 2004MNRAS.352..535D}.

In accordance with the spherical collapse model the signal extends out
to  $\sim   2\Rvir$,    which  roughly   reflects   the   distance  of
turnaround.  At $2.5\Rvir$  the  distribution  is close  to  isotropic
suggesting that at  these distances the impact  of the group potential
is   negligible and the  satellite  motions  are   dominated by  local
potential variations arising from the  filaments and dark matter halos
within these filaments.  Note  that the  presence of this  filamentary
structure  in   the   vicinity  of  groups is   clearly  evident  from
Figure~\ref{fig:a11}.   Finally, the  increase  of the radial velocity
alignment on large scales,   $\gtrsim4\Rvir$,  is simply due   to  the
Hubble flow (i.e., $\vel
\rightarrow  1$  at $r  \rightarrow  \infty$).   The  middle panel  of
Figure~\ref{fig:vel} shows $\vel$ for the inward moving satellites only.
The radial trend within $2\Rvir$  is somewhat enhanced compared to the
upper  panel. At  larger radii,  the inward  moving satellites  have a
velocity structure that is  consistent with isotropy.  The lower panel
of Figure~\ref{fig:vel}  reveals a marked difference in  the behavior of
$\vel$  for the outward  moving satellites.   It indicates  a slightly
radial trend for  satellites within $1\Rvir$ which is  much lower than
seen in the  upper two panels.  Within $1-2\Rvir$  it drops below 0.5,
indicating a preference for  tangential velocities.  Together with the
information  derived  from  Figure~\ref{fig:io}  this  suggests  that  a
substantial   fraction  of  outward   moving  satellites   located  at
$1-2\Rvir$  currently  are close  to  their  apo-center passage  after
having  crossed the  more central  regions of  the group.  Finally, on
large scales  the outward moving satellites clearly  reveal the Hubble
flow.

\begin{figure}
\plotone{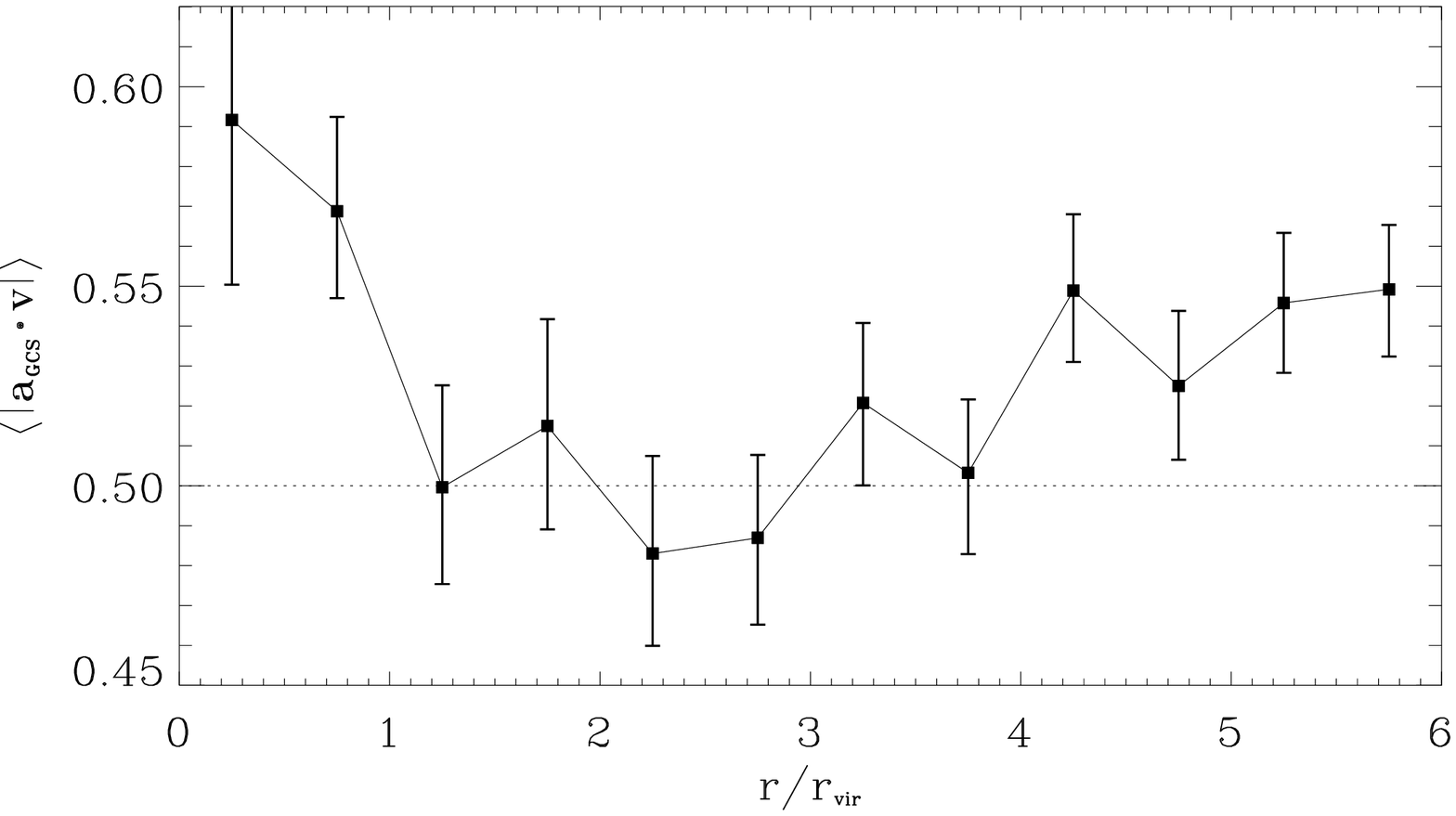}
\caption{\label{fig:c1v} Same as Figure~\ref{fig:a11},  but for $\cIv$.}
\end{figure}
Figure~\ref{fig:c1v} displays the  radial dependence  of $\cIv$  which
measures  the cosines of the angels   between the satellite velocities
and the  orientation of the GCS.  On  large  scales the radial outward
motion  caused by the Hubble flow  exceeds the  internal velocities of
the satellites  within  the filaments.   Since   the GCS is   strongly
aligned  with  these  filaments over  the  entire  radial range  shown
(cf. Figure~\ref{fig:a11}), one has that  $\cIv > 0.5$ on scales where
the Hubble flow    becomes important ($\gtrsim4\Rvir$).    The  strong
alignment signal on small scales indicates that the satellites tend to
move parallel to the orientation of the GCS.  According to
\cite{1997MNRAS.290..411T} and \cite{2006MNRAS.367.1781A} the
principal    axes of  the   velocity  anisotropy  tensor are  strongly
correlated  with     the    principal    axes of      the    satellite
distribution.   Therefore,   the     alignment   found    for   $\aII$
(Figure~\ref{fig:a11})  actually   implies  an  analogous   signal for
$\cIv$.   However, in  contrast to  the  halo  alignment, $\aII$,  the
velocity    halo   alignment,      $\cIv$,  only  extends     out   to
$\sim1\Rvir$.   Beyond this  radius   a  substantial  fraction of  the
satellites shows relatively large angles  between their velocities and
the orientation of  the GCS which is  consistent  with the picture  of
tangential motions associated    with the apo-center  passage   of the
satellites, as discussed in the context of Figure~\ref{fig:vel}.

\begin{figure}
\plotone{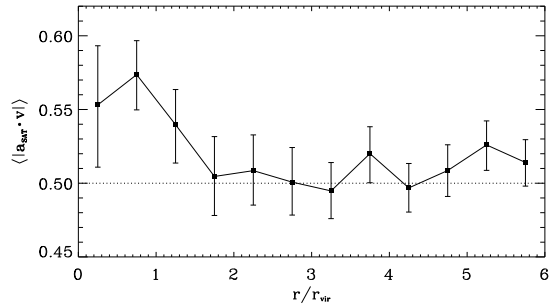}
\caption{\label{fig:a1v} Same as Figure~\ref{fig:a11},  but for the
distributions    of cosines   between   the  satellites  velocities and
positions, $|\vec{v}\!\cdot\!\vec{r}|$.}
\end{figure}
Finally  we  consider  the  auto  velocity  alignment,  $\aIv$,  which
reflects  the  distribution  of  the  cosines  between  the  satellite
velocities   and   their   orientations,  $|\vec{v}\cdot\vec{a}_{\rm SAT}|$.   
Fig~\ref{fig:a1v} displays the variation of $\aIv$ with $r/\Rvir$.  

The   signal for  $\aIv$ shows  a   maximum at  $0.7\Rvir$.  At larger
distances  it decreases quickly.   Beyond $1.5\Rvir$  it is roughly in
agreement with an isotropic  distribution.  A possible reason  for the
slight central dip  is, that satellites  on their peri-center passages
move  perpendicular  to   the  gradients   of  the   group  potential.
Figure~\ref{fig:ax1},   however, revealed    a preferential     radial
orientation of   these   satellites.  Thus,   during   the peri-center
passages the angles between satellite  orientations and velocities can
become large. The degree of the radial  alignment depends on the ratio
between the internal  dynamical time of  the satellite,  with which it
can adjust  its   orientation, and the   duration  of  the peri-center
passage.   If the peri-center passage occurs  too quickly the time may
be too short for a `perfect' radial alignment
\citep[cf.,][]{2007arXiv0705.2037K}.  On  large  scales ($1-2\Rvir$) a
similar mechanism may  take place.  Above we  have  argued that within
this distance range a substantial fraction of  satellites are close to
their apo-center passage.  During this phase  the velocities are again
perpendicular to the  gradient of the  potential but, as indicated  by
Figure~\ref{fig:ax1}, the  satellites   are  oriented radially.    The
comparison between the signal for  $\axI$ and $\aIv$ suggests that, in
a statistical sense,  the (spatial)   radial alignment is   maintained
during the entire orbit of the satellite  within the potential well of
the groups, which in turn causes a suppression of  $\aIv$, at its apo-
and peri-center.

\section{Projected Alignments}
\label{sec:2D}
To facilitate  a comparison with  observations, in particular with the
results presented  in Paper~I, we  repeat the foregoing analysis using
projected data, i.e. we project the particle distribution along one of
the   coordinate axes and  compute the  second moment of  mass for the
projected particle distribution. Accordingly,
for the  distances between  GCS and  satellites we  use  the projected
values (all satellites within a  sphere of $6\Rvir$  about the GCS are
projected), which we label as $R$ (the physical distances are labelled as
$r$).

Since   the  projections   along  the  three   Cartesian
coordinate axes are independent we   include all three projections  of
each host-satellite  in our 2D sample.   To reduce the contamination by
satellites   associated with massive  ambient  groups we exclude those
host-satellite systems where another  SKID group more massive than the
GCS (which  is  most likely the  center  of an ambient  host-satellite
system) is found  within a  sphere of  $6\Rvir$.   After rejection  of
`contaminated' groups   we  obtain 1034  and 543  satellites   for the
$1.0r_\sigma$ and $0.5r_\sigma$ samples with  3D distances to the  GCS
$\leq\Rvir$ (for all groups irrespective of their environment we found
1431  and 772,   see  \S~\ref{sec:ori}.)  Furthermore, since   (due to
technical reasons) we project satellites located within a {\it sphere}
of $6\Rvir$ the projected volume at  large projected distances shrinks
substantially.  Therefore,  we analyze the  2D data  only for projected
distances $\lesssim3\Rvir$ which roughly  resembles the  projection of
all satellites within a cylinder with a radius  $3\Rvir$ and length of
$10\Rvir$ along the `line of sight'.  Thus,  in an approximate manner,
uncertainties in the  determination   of group membership   based on
redshift measurements are accounted for.

The resolution of  the simulation does not  permit to  probe alignment
below $0.3\Rvir$.      Other   authors  \citep[using   semi-analytical
techniques, e.g.,][]{2007MNRAS.378.1531K}  have  bypassed this problem
by introducing so-called {\it   orphan galaxies}, i.e.  galaxies which
are associated  with the once most bound  particle of a satellite halo
which subsequently has  become undetectable due   to the stripping  by
tidal forces.  Here we do  not adopt this technique  since it does not
provide us  with  information about  the  orientation of a  satellite.
Both approaches,  considering  only satellite  halos  with  a  minimum
number of  particles and  the introduction  of  orphan  galaxies, have
certain disadvantages.  The former does not account for galaxies which
are hosted by strongly  stripped subhalos  whereas the latter  ignores
the dynamical differences of galaxies and (once most bound) particles.

The  application of a fixed  lower  particle limit excludes satellites
from  the analysis which   still    constitute distinct objects.    In
particular satellites which  are strongly  tidally stripped  may  fall
below the selection criterion even if  the galaxy, which is assumed to
sit at the center, may still be observable.  Thus, we caution that our
satellite sample may be somewhat  biased toward more recently accreted
satellites  compared to a hypothetical  galaxy population. This effect
appears whenever a  fixed lower   particle  limit is imposed.

\begin{figure}
\plotone{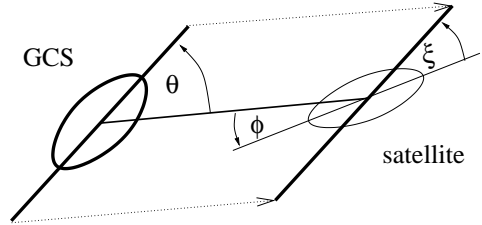}
\caption{\label{fig:sketch}
Illustration of the three angles $\theta$, $\phi$ and $\xi$, which are
used  for halo  alignment,  radial  alignment  and  direct  alignment,
respectively \citep[cf.,][]{2007ApJ...662L..71F}.}
\end{figure}
In analogy to Paper~I we define  the angles $\theta$, $\phi$ and $\xi$
to  address the projected  halo,  radial and  direct  alignments (same
definitions    as   in \S\ref{sec:3D}  but   for   the   2D data,  see
Fig.~\ref{fig:sketch}) and  the projected orientations are referred to
as position angles  (PAs). It is not  straightforward  to derive galaxy
properties, such    as luminosity and   color, from  the   dark matter
distribution.  In particular, if the satellite  halo hosts a late type
galaxy, it is not obvious  how to accurately determine the orientation
of the disk (but see e.g., \citealt{2007MNRAS.378.1531K} and
\citealt{2007arXiv0704.3441A} for attempts). 
On  the    other side, if  one   focuses  on early type   galaxies the
orientation  of  the central dark matter   distribution is very likely
correlated  with the orientation  of   the stellar component (see  the
evidence from gravitational lensing, e.g.,
\citealt{2002sgdh.conf...62K}).  The   lower  particle  limit  for the
satellites  results  in   a   lower mass   of $10^{11}\hMsol$   within
$25\hkpc$.    Assuming  a  dynamical mass-to-light    ratio of  a  few
\citep[][]{2006MNRAS.366.1126C}  within  this radius  yields a stellar
component  which roughly resembles  $L_\ast$  galaxies. Therefore, our
findings in the current paper may be  best compared with results based
on bright  early-type satellite galaxies.  However, as we have pointed
out   in   Paper~I, our observational   results   were only marginally
dependent  on the luminosity/mass  of satellite galaxies. Therefore, a
comparison with observations  based on somewhat fainter  satellites is
viable as well.
\subsection{Halo alignment}
\begin{figure}
\plotone{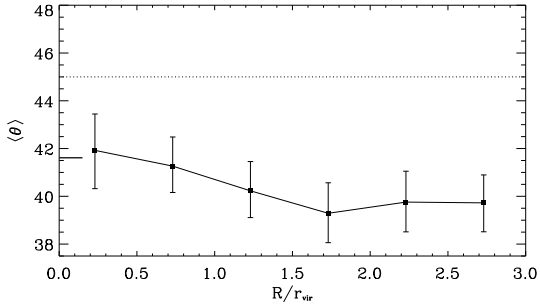}
\caption{\label{fig:D2a11}
Mean   angle,  $\theta$,  between the   PA of  the    GCS and the line
connecting the GCS   and  a satellite,   as a  function of   projected
distance $R/\Rvir$  with equidistant  bins  of  $0.5\times\Rvir$.  The
error bars give the $95\%$ bootstrap confidence intervals for the mean
angles within   each  bin. The  short  horizontal   line on the   left
indicates the signal for the innermost bin if only the satellites with
in 3D  distances  $\leq1\Rvir$ are   projected. The corresponding   3D
results are shown in Fig.~\ref{fig:a11}.}
\end{figure}

Figure~\ref{fig:D2a11}  shows   the  results  obtained  for  the angle
$\theta$ between  the orientation of the  GCS  and the line connecting
the GCS  with the satellite.  The   short horizontal line on  the left
indicates  the  result for the innermost  bin  if only  the satellites
within $1\Rvir$ are projected.  The sample shows $\langle\theta
\rangle < 45^{\circ}$ for the entire distance range. The
error bars give the $95\%$ bootstrap confidence intervals for the mean
angles  within  each  bin. The alignment  strength  within  $\Rvir$ is
$\sim42^\circ$, in good agreement with the findings of
\cite{2005ApJ...628L.101B},  \cite{2006MNRAS.369.1293Y}. In Paper~I we
found a mean value $\theta\approx41^\circ$  within $0.5\Rvir$ which is
very close  to   the  values we   obtain  for the  innermost  bin,  in
particular if only the   satellites within $1\Rvir$ (short  horizontal
lines on the left) are projected. As also shown by
\cite{2006ApJ...650..550A} the alignment signal extends beyond the
virial radius.  The strongest amplitude  is  found outside the  virial
radius   at  $\sim   1.7\Rvir$.   Currently  there  are  no  available
observations  covering  the  same  distance  range.   The  analysis in
Paper~I,  for instance, is based  on galaxies within the virial radius
whereas      we   use  all   galaxies       with  projected  distances
$\lesssim3\Rvir$. According to our findings a  search for alignment of
satellite distribution in group environments for distances larger than
$\Rvir$ may be a promising proposition. 
\subsection{Radial alignment}
\begin{figure}
\plotone{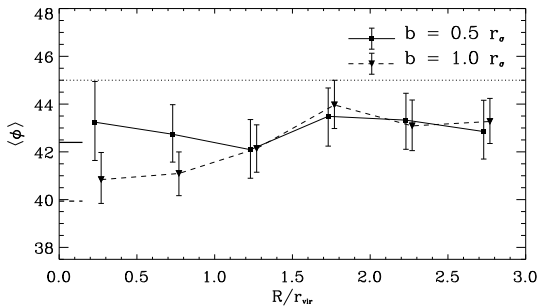}
\caption{\label{fig:D2ax1}
Same as Figure~\ref{fig:D2a11}, but for the angle $\phi$. In addition
the radial dependence  of  the $b=1 r_\sigma$  sample is  displayed as
well. The corresponding 3D results are shown in Fig.~\ref{fig:ax1}.}
\end{figure}
Figure~\ref{fig:D2ax1}  displays the mean  angle $\phi$ between the PA
of  the  satellite  and the  line   connecting the satellite  with its
GCS. For all group centric distances there  is a clear and significant
signal for the  major axes of the satellites  to point towards the GCS
(i.e., $\langle\phi\rangle < 45^\circ$).  The projection of only those
satellites  within $1\Rvir$  increases   the central signal  by  about
$1^\circ$ (differences between the innermost data points and the short
horizontal lines). The mean angle for  the $0.5r_\sigma$ sample within
the innermost bin is $\sim43^\circ$  and according to Paper~I the mean
value for the red  SDSS satellites within  $0.5\Rvir$ is very close to
this value. However, the  observations suggest a significant alignment
for  red  galaxies  only  out to  $0.7\Rvir$  whereas  the N-body data
indicate that radial     alignment  extends  beyond $3\Rvir$.      The
discrepancy may be caused by the observational confinement to galaxies
within the virial radius.
\subsection{Direct alignment}
\begin{figure}
\plotone{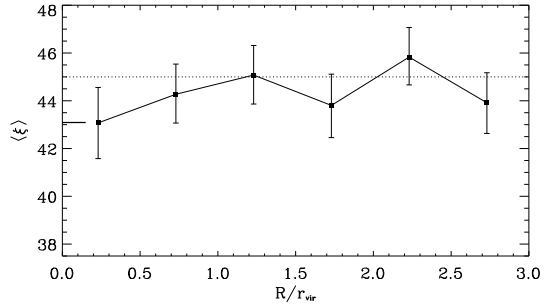}
\caption{\label{fig:D2ac1}
Same  as Figure~\ref{fig:D2a11},  but  for the  angle $\xi$.}
\end{figure}
Figure~\ref{fig:D2ac1} displays  the results for the direct alignment,
based  on  the  angle  $\xi$ between  the   orientations  of GCSs  and
satellites.  The  alignment signal  is significant at  a $\gtrsim95\%$
confidence  level  for  distances  $\lesssim0.5\Rvir$.  In Paper~I  we
obtained $\xi\approx44^\circ$ for  red  satellite with in   $0.5\Rvir$
which indicates a somewhat weaker alignment  than we find here.  Since
the  3D analysis   shows  no   increase  of  $\aII$  at small   scales
(Figure~\ref{fig:a11}) the  central enhancement displayed  here has to
be interpreted as a result of projection effects.
 
In  summary for all  three types of  alignments we find good agreement
between numerical data   presented here and  the observational results
from Paper~I.  In particular the relative strength among the different
alignments is  well reproduced  in   the numerical  analysis.  Due  to
limited resolution the range  below $1\Rvir$ is only  sparsely sampled
thus   no  detailed information about the    radial  dependence of the
alignment signal  on small scales can  be derived. However, the signal
for $\theta$ increases with distance  which is only marginally implied
by the  SDSS   results presented  in  Paper~I.   Also  for $\phi$, the
	 dependence  on   the    distance  disagrees between   simulations
and observations. It  is   currently  unclear   whether this  is   due
		 to shortcomings from the numerical or observational side.
\section{Summary}
\label{sec:diss}
Based  on   a  sample   of  515  groups   with  masses   ranging  from
$10^{13}\hMsol$  to $5\times10^{14}\hMsol$  we  have investigated  the
halo alignment,  $\aII$, the radial  alignment, $\axI$ and  the direct
alignment $\acI$, between  the central region of each  group (the GCS)
and  its  satellite  halos  out  to a  distance  of  $6\Rvir$.   Here
$\vec{a}_{\rm GCS}$,  $\vec{a}_{\rm SAT}$  and  $\vec{r}$ denote  the  unit  vectors
associated with  the orientation  of the GCS,  the satellites  and the
line  connecting both  of them.   Additionally, we  have  employed the
directions  of  the  satellite   velocities  $\vec{v}$  to  probe  the
alignments $\vel$, $\cIv$ and $\aIv$,  referred to as radial, halo and
auto velocity alignments, respectively. Our main results are:
\begin{itemize}
\item[(1)] Halo, radial and  direct alignment differ in strength.  The
  halo alignment is strongest followed by the radial alignment. By far
  the   weakest  and least significant  signal  comes  from the direct
  alignment.  This sequence is found in the 3D analysis as well as for
  the projected  data and  agrees well   with our  recent analysis  of
  galaxy alignments in the SDSS (cf., Paper~I).
\item[(2)] The signal  for the halo alignment, $\aII$, reaches  far 
  beyond the virial radii of the groups ($>6\Rvir$) which we interpret
  as evidence for large scale filamentary structure.
\item[(3)] The signal  for the radial alignment, $\axI$, is  largest
  on small scales.  After a rapid  decline with distance it  flattens,
  such that a   relatively small  $\axI\approx0.52$,  but  significant
  deviation from isotropy  is  detected out to $\sim  6\Rvir$. Whereas
  the small scale signal more likely  owes to the group's tidal field,
  the weak   but significant  signal  on large   scales suggests  that
  satellites tend to  be oriented  along the  filaments in which  they
  reside.
\item[(4)] The 3D signal for the direct alignment, $\acI$, shows a
  weak trend for   parallel orientations on   scales $\lesssim2\Rvir$.
  The   projected data indicate an   increasing  signal for  distances
  $\lesssim0.5\Rvir$  which is likely   caused  by  projection effects. 
\item[(5)]  All  kinetic alignment  signals  are  highly
  significant  at small scales.   The  signal for $\vel$  is basically
  constant within  $2.0\Rvir$, beyond which it  rapidly drops.  In the
  subset     of  outward moving satellites    we  find  a tendency for
  tangential motions which can be  attributed to the satellites  which
  have been accreted earlier and are currently  passing their peri- or
  apo-centers.  The signal for $\cIv$  is maximal at the center, drops
  rapidly with distance  and disappears at $1\Rvir$.  Finally,  $\aIv$
  shows a slight  dip at the  center, reaches a maximum at $0.7\Rvir$,
  and  becomes consistent  with    isotropy at $1.5\Rvir$.  All  these
  features  support the    interpretations advocated  for the  spatial
  alignments.
\end{itemize}
The simulation analyzed here  clearly  demonstrates that tidal  forces
cause a      variety of  alignments  among    neighboring, non-linear
structures.   On  large scales, the   tidal forces are responsible for
creating a  filamentary network, which  gives rise to a halo alignment
out    to at least $6\Rvir$.   The   same tidal  forces  also cause an
alignment between filaments and (sub)structures within the filaments
\citep[cf.,][]{2006MNRAS.370.1422A,2007arXiv0704.2595H}
which in turn   results in a  large  scale radial alignment  with  the
virialized structures at  the nodes of  the cosmic  web.  Within these
virialized structures,    tidal forces are  responsible for   a radial
alignment of its substructures, similar to the tidal locking mechanism
that affects the Earth-Moon system.  This is  further supported by the
fact that the auto  velocity alignment $\aIv$ reveals  a dip  on small
scale, indicating that at  peri-centric passage satellites tend  to be
oriented    perpendicular   to   the    direction  of    their  motion
\citep[cf.,][]{2007arXiv0705.2037K}.   This behavior also   explains,
why the direct   spatial alignment, $\acI$,  is so  weak.   A possible
direct alignment    originating  from the co-evolution    of group and
satellites, as  proposed   by \cite{2005ApJ...629L...5L}, is   quickly
erased  as the satellites  orbit in the  potential well  of the group.
For  future   work it  will be  instructive   to  trace  the orbits of
individual  satellites and consider more  closely how their shapes and
orientations evolve with time.

The infall regions around virialized dark matter halos cause a radial
velocity alignment out to $\sim 2\Rvir$, and  an enhancement of inward
moving (sub)structures.  At around the same scale, the (sub)structures
with  a  net outward  movement  have a tendency to  move tangentially.
This most  likely reflects the  apo-centric passage  of substructures
that  have  previously fallen through  the  virialized halo.  Within a
virialized region, the orientation of orbits is naturally aligned with
that of its GCS. Since (sub)structures reveal  at most a weak velocity
bias with respect to dark matter particles
\citep[e.g.,][]{2006MNRAS.369.1698F},  this   causes  a  strong    halo
velocity alignment on scales $\la \Rvir$.  The halo velocity alignment
is   also strong on large  scales  ($\ga 3\Rvir$),  which reflects the
Hubble flow combined  with the filamentary, non-isotropic distribution
of (sub)structures on these scales.  

A one-to-one comparison between the N-body  results discussed here and
the    observations    presented    in   Paper~I is not
straightforward. Although we have employed the same  mass range for the groups
in both studies  the resolution of the  current simulation only allows
to  resolve  satellites which are   expected to  host $\gtrsim L_\ast$
galaxies.  These are  bright compared to  our SDSS sample for  which a
lower magnitude  limit of  $^{0.1}M_r -  5\log h  \leq -19$  has  been
adopted.  Nevertheless, the qualitative agreement between the relative
strengths of   the   different   types   of   spatial alignment     is
promising. Supplementary  to  the observational results of  Paper~I we
find a strong halo  alignment and a  somewhat weaker  radial alignment
out to at least $6\Rvir$ which we will investigate further.

Finally, the weak but significant detection of radial alignment out to
$6\Rvir$ may  contaminate  the  cosmic  shear measurements   on  these
scales.   This  correlation has to   be  considered, either by  simply
removing  or  down-weighting pairs  of  galaxies  within this distance
range \citep{2002A&A...396..411K, 2003MNRAS.339..711H}. This may be 
particularly important for applications of weak gravitational lensing
for the purposes of precision cosmology.
\acknowledgements
This work has is  supported by NSFC (10533030, 0742961001, 0742951001,
973 Program No. 2007CB815402) and  the Knowledge Innovation  Program of
the  Chinese  Academy of  Sciences,   Grant No. KJCX2-YW-T05.  The CAS
Research  Fellowship  for  International Young  Researchers  (AF), the
local support of the Chinese Academy of  Sciences (HJM and SM) and the
Alexander von  Humboldt Foundation  (SM)  is gratefully  acknowledged.
HJM would like  to acknowledge the  support  of NSF ATP-0607535,  NASA
AISR-126270, and NSF IIS-0611948 .

\end{document}